\documentclass[%
 reprint,
 amsmath,amssymb,
 aps,
pra,
superscriptaddress]{revtex4-2}

\usepackage{graphicx}
\usepackage{dcolumn}
\usepackage{bm}

\begin{document}

\preprint{APS/123-QED}

\title{Vertical engineering for large stimulated Brillouin scattering in unreleased silicon-based waveguides}

\author{Laura Mercad\'e}
\affiliation{Nanophotonics Technology Center, Universitat Polit\`ecnica de Val\`encia, Camino de Vera s/n, 46022 Valencia, Spain}
\author{Alexander V. Korovin}
\affiliation{Institute for Physics of Semiconductors, NAS of Ukraine, Prospect Nauki, 41, Kiev, 03028, Ukraine}
\affiliation{Institut d'\'Electronique, de Micro\'electronique et de Nanotechnologie,  UMR CNRS 8520, Universit\'e de Lille, Cit\'e Scientifique, 59655 Villeneuve d'Ascq, France}
\author{Yan Pennec}
\affiliation{Institut d'\'Electronique, de Micro\'electronique et de Nanotechnologie,  UMR CNRS 8520, Universit\'e de Lille, Cit\'e Scientifique, 59655 Villeneuve d'Ascq, France}
\author{Bahram Djafari Rouhani}
\affiliation{Institut d'\'Electronique, de Micro\'electronique et de Nanotechnologie,  UMR CNRS 8520, Universit\'e de Lille, Cit\'e Scientifique, 59655 Villeneuve d'Ascq, France}
\author{Alejandro Mart\'inez}
\affiliation{Nanophotonics Technology Center, Universitat Polit\`ecnica de Val\`encia, Camino de Vera s/n, 46022 Valencia, Spain}

%
%
%


\date{\today}

\begin{abstract}
Strong acousto-optic interaction in silicon-based waveguides generally requires releasing of the silicon core to avoid mechanical leakage into the underlying silica substrate. This complicates fabrication, limits thermalization, reduces the mechanical robustness and hinders large area optomechanical devices on a single chip. Here, we overcome this limitation by employing vertical photonic-phononic engineering. Specifically, the insertion of a thick silicon nitride layer between the silicon guiding core and the silica substrate contributes to reduce GHz-frequencies phonon leakage enabling large values of the Brillouin gain in an unreleased platform. We get values of the Brillouin gain around 300 (Wm)$^{-1}$ for different configurations, which could be further increased by operation at cryogenic temperatures. These values should enable to observe Brillouin-related phenomena in cm-scale waveguides or in more compact ring resonators. This finding could pave the way towards large-area unreleased cavity and waveguide optomechanics on silicon chips.
\end{abstract}

\maketitle


\section{\label{Intro}Introduction}

Stimulated Brillouin Scattering (SBS) is a nonlinear process arising from photon-phonon interaction in material systems supporting both optical and acoustic propagating waves \cite{BRI22-AP,CHI64-PRL}. For many years, optical fibers were the most extended platform to observe SBS and related phenomena \cite{IPP72-APL}. Indeed, many applications - mainly in the context of microwave photonics \cite{CAP07-NP}- were developed, ranging from tunable microwave filters \cite{VI07-OL} to optical pulse storage \cite{ZHU07-SCI}. The current trend towards miniaturization and on-chip integration of optical devices has also been followed for the case of SBS-related phenomena and applications \cite{EGG19-NP}. The first demonstration of on-chip SBS was made by using a chalcogenide glass ($Al_{2}S_{3}$) as guiding material for both light and sound \cite{PANT11-OE}. As in optical fibers, acousto-optic (AO) interaction via electrostriction and photoelasticity in the waveguide core material was the main responsible for SBS \cite{PEN14-NP}. Despite that this approach enables large Brillouin gain and multiple SBS-based on-chip functionalities \cite{CHO17-JLT,MAR15-OPT}, it is not compatible with CMOS technology, which prevents massive integration with mainstream silicon photonic and electronic components.

Silicon was first disregarded as an appropriate material for integrated SBS since when placed over silica – as in standard silicon-on-insulator wafers - it cannot confine mechanical waves, which prevents strong AO coupling \cite{SAF19-OPT}. SBS in silicon waveguides requires the waveguide to be at least partly released so that phonon pathways towards the silica substrate are strongly inhibited \cite{SHIN13-NCOMM,LAER15-NP,LAER15-NJP}. Noticeably, the strong field confinement in wavelength-size silicon core does not only result in enhanced AO volumetric effects; in addition, the photon-phonon interaction is mediated by the radiation pressure exerted by the optical field on the interfaces, which scales up with the electric permittivity contrast at the boundary \cite{PEN14-NP,CHAN12-APL}. However, besides recent exciting demonstrations \cite{OTT18-SCI}, the need for partial release of the silicon core complicates the fabrication (it requires HF baths), reduces thermal dissipation, prevents the introduction of more sophisticated AO elements (required for building chips-scale processors based on SBS \cite{EGG19-NP}) and reduces the mechanical robustness of the device. Therefore, a silicon-compatible platform with mechanical robustness, high Brillouin gain and low propagation losses, which does not require for releasing the silicon core, is still missing \cite{EGG19-NP}. 

Recently, a novel approach for unreleased-silicon SBS based on a geometric softening technique was proposed \cite{SAR16-APLP}. Here, two high aspect ratio silicon waveguides spaced by a thin slot and placed on a silica substrate were shown to provide large values of the Brillouin gain ($G_{B}$) for both backward and forward configurations. However, fabrication of this slotted structure becomes complicated because of the thin slot and optical propagation losses may become quite large because of the large area of the sidewalls boundaries, so SBS has not been demonstrated in this system to the best of our knowledge. Silicon nitride waveguides have also been shown as good candidates for unreleased SBS-based devices mainly because they show extremely low linear losses and negligible two-photon absorption \cite{GUN19-NP}. In this case, phonons are not confined in the waveguide core, which results in $G_{B}$ values that are orders of magnitude smaller than in released silicon waveguide \cite{GYG20-PRL} and, consequently high operating optical power and ultralong waveguides are required.

\begin{figure*}[t]
\center
\includegraphics[width=\textwidth]{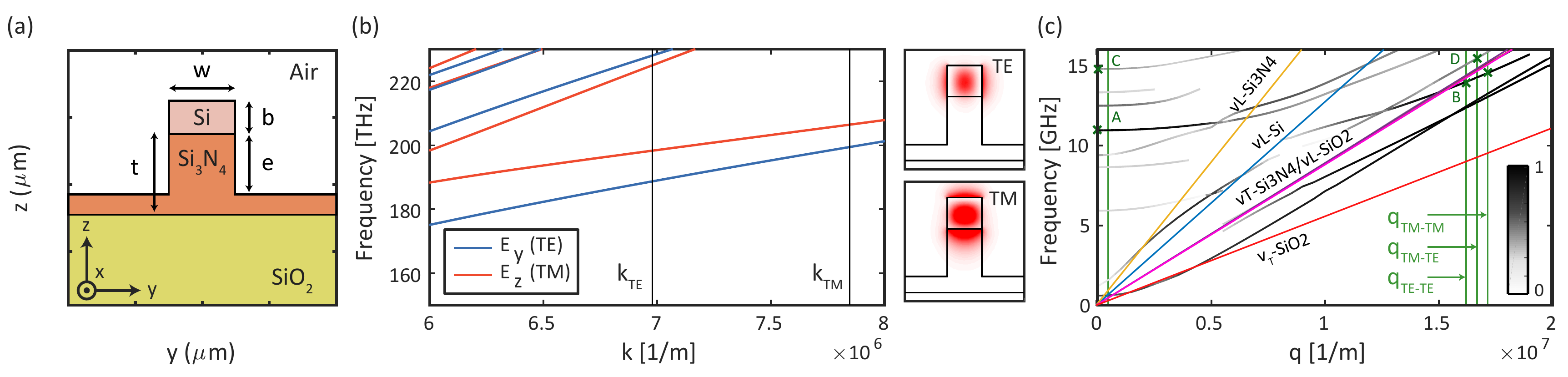}
\caption{$Si$-$Si_{3}N_{4}$ photonic-phononic waveguide. (a) Material cross-section of the waveguiding structure; (b) Calculated dispersion relation of the TE-like and TM-like optical  guided modes Ey (TE-like mode) and Ez(TM-like mode) with the optical wavevectors involved for each mode. Insets: Computed Ey and Ez components of the optical TE-like and TM-like waveguide modes, respectively. (c) Calculated dispersion relation for the mechanical modes.  The interfaces between $Si/Si_{3}N_{4}$ and air are treated as free boundaries and perfectly matched layer conditions are imposed at the substrate limits interfaces. Here, the colorscale shows a localization factor of the mechanical mode within the silicon core, so that higher localization should in principle provide higher SBS gain. The optical and mechanical wave vectors involved in some of the calculations below are also depicted in (b) and (c), respectively. The dimensions of the waveguide in (b) and (c) are w=325 nm, b=0.9w, e=450 nm and t=600 nm.}
\label{fig:intro}
\end{figure*}

In this work, we introduce a novel unreleased lithographically-defined silicon-based waveguide that can confine both optical and mechanical waves in the silicon core and provide large $G_{B}$ values. We engineer the waveguide vertically by inserting a thick silicon nitride layer between the top silicon waveguide and the silica substrate. This enormously reduces phonon leakage and enables phonon propagation below the sound line at sufficiently large wave vectors $q$, leading to efficient ($G_{B}\simeq$ 300 1/Wm) forward and backward stimulated Brillouin scattering at room temperature.Moreover, this total SBS gain can be enhanced up to 10$^{5}$ at low temperatures. From the fabrication point of view, the whole structure can be fabricated using standard lithography (either electronic or optical) and etching, not requiring partial or full releasing via HF baths.

\section{\label{sec:level1}Description of the photonic - phononic waveguiding system}

The key idea of our AO waveguiding system is to place between the guiding layer (silicon) and the substrate (silica) an intermediate layer made of a material having both a refractive index lower than silicon (to ensure photon confinement in the silicon layer) and a sound speed higher than silicon (to avoid phonon leakage from silicon to silica). We choose silicon nitride ($Si_{3}N_{4}$) to build this intermediate layer, which is completely compatible with silicon photonics technology. Noticeably, even though it can depend on its state (crystalline or amorphous) as well as the deposition method, the speed of sound in $Si_{3}N_{4}$ is higher than in silicon \cite{GYG20-PRL,WOL14-OE,LIUARX-17}, which is a basic requirement to reduce mechanical leakage from the silicon layer to the silica substrate.

Figure \ref{fig:intro}(a) shows a transverse cross section of the proposed waveguiding system: a silicon ($Si$) core (width $w$, thickness $b$) lies upon an intermediate $Si_{3}N_{4}$ layer (width $w$, total thickness $t$, etched thickness $e$) which is placed over a silica substrate, which we assume to be semi-infinite. For the numerical simulations we assume that the refractive indices of the involved materials at telecom wavelengths are $n_{Si}$=3.48, $n_{SiO_{2}}$=1.44 and $n_{Si_{3}N_{4}}$=2. By properly choosing the dimensions, the waveguiding system supports a set of optical guided modes (namely, the TE-like and TM-like modes) in the relevant telecom band, as depicted on Fig. \ref{fig:intro}(b). To perform mechanical simulations, we choose the following parameters. The photoelastic constants in contracted notation are [$p_{11}$=-0.09, $p_{12}$=0.017,$p_{44}$=-0.051] for silicon \cite{QIU13-OE}, which is also assumed to be an anisotropic material with a density $\rho$=2329 [kg/m$^{3}$] and stiffness constants in Voigt notation of [$c_{11}$=2.17$\cdot$10$^{11}$, $c_{12}$=4.83$\cdot$10$^{10}$, $c_{44}$=6.71$\cdot$10$^{10}$ [Pa] \cite{SAY16-OE,AUL73-BOOK}. For the silica substrate we consider photoelastic constants of [$p_{11}$=0.121, $p_{12}$=0.27, $p_{44}$=-0.075] \cite{SAR17-OPT} whilst for the mechanical properties we used a Young’s modulus = 73.1 [Pa], a Poisson’s ratio of $\nu$=0.17 and a density $\rho$=2203 [kg/m$^{3}$]. Finally, $Si_{3}N_{4}$ was considered to be also an anisotropic material with $\rho$= 3100 [kg/m$^{3}$] and stiffness constants of [$c_{11}$=4.33$\cdot$10$^{11}$, $c_{33}$=5.74$\cdot$10$^{11}$, $c_{44}$=5.74$\cdot$10$^{11}$, $c_{66}$=1.19$\cdot$10$^{11}$, $c_{12}$=1.95$\cdot$10$^{11}$, $c_{13}$=1.25$\cdot$10$^{11}$] [Pa] \cite{VOG00-APL}. Notice that the photoelastic coefficients of the $Si_{3}N_{4}$ are assumed to be equal to those of silica as they have been largely unknown in the literature \cite{JIN18-OE} though a recent experiment has reported an experimental value $p_{12}$= 0.047 \cite{GYG20-PRL}. With these mechanical parameters and the previous dimensions of the waveguide, we obtain a set of mechanical modes as show in Fig. \ref{fig:intro}(c), where $q$ is the wavevector along the axis of the waveguide. More details regarding the mechanical simulations, including the effect of the area of the substrate considered in the calculations, can be found in the Supplementary Material.

Notice that the mechanical dispersion relation shows all the modes that we can find in the structure and the color scale represents the acoustic localization factor in the silicon core, calculated as ($\int_{S_1}dS |\mathbf{u}|^2)/(\int_{S_2}dS |\mathbf{u}|^2$) where $u$ is the total mechanical mode displacement, and $S1$ and $S2$ denotes the silicon region and the full cross-section under consideration respectively. Darker lines represent a localization close to unity, which means that the whole mechanical mode is located in the silicon region. Meanwhile, grey lines show less localized modes. In Fig. \ref{fig:intro}(c) the mechanical  modes with the highest SBS gain (see below) for different intramodal and intermodal configurations have been highlighted. The mechanical modes termed A and C give the highest $G_{B}$ values in the case of forward intramodal TE and TM coupling, respectively. The mechanical mode B is involved in both backward intramodal TE-TE and TM-TM interaction whilst the mechanical mode D is the one that gives a higher $G_{B}$ value in the backward intermodal TM-TE configuration.

\section{\label{sec:level1}AO interaction and Brillouin gain}

The optical and mechanical modes will interact via volumetric and surface processes as long as energy and momentum are conserved ]\cite{EGG19-NP,PEN14-NP,SAF19-OPT}. In an AO waveguiding system as the one presented here, $G_{B}$ for a certain mechanical mode can be obtained as \cite{QIU13-OE}


\begin{eqnarray}
G_{B}= \frac{2\omega Q_{m}}{\Omega_{m}^{2}}\frac{\vert\langle \textbf{f},\textbf{u}_{m}\rangle\vert^{2}}{\langle P_{p}\rangle\langle P_{s}\rangle\langle \textbf{u}_{m},\rho \textbf{u}_{m}\rangle}
\label{eq:sbs1}
\end{eqnarray}

\noindent where $m$ is the mechanical mode number (omitted in what follows for the sake of simplicity), $\langle u_{m},\rho u_{m}\rangle$ is the mechanical mode effective mass, $\langle P\rangle = \frac{1}{2}v_{g}\langle \textbf{E},\epsilon \textbf{E}\rangle$ is the averaged Poynting vector for guided pump (``$p$'') and Stokes (``$s$'') waves, $\textbf{E}=(\textbf{E}_{p}e^{i(-k_{p} x+\omega_{p}t)}+\textbf{E}_{s}e^{i(-k_{s} x+\omega_{s}t)})/2 + c.c.)$ is the electric field, $\textbf{D}=\epsilon \textbf{E}$ is the displacement field, $\epsilon$ is the material permittivity, $v_{g}$ is the group velocity, $\textbf{f}$ are the involved optical forces, $\textbf{u}=u_{x}\textbf{x}+u_{y}\textbf{y}+u_{z}\textbf{z}$ is the elastic displacement caused by the total optical power created by the pump and Stokes waves, and $\Omega_{m}$ and $Q_{m}$ are the frequency and the mechanical quality factor of the involved mechanical modes at $q=k_{p}-k_{s}$. Notice that in all our calculations we have assumed a mechanical Q factor $Q_{m}$ = 10$^{3}$, meaning that the mechanical damping is limited by material losses \cite{MAC19-ARX}. This is a typical convention followed in cavity and waveguide optomechanics at GHz frequencies when operated at room temperature. However, in our calculations below we have kept the original mechanical quality factor for those modes showing $Q_{m}$ $<$ 10$^{3}$ in simulations, in order to not to overestimate leaky modes. In the last part of our paper, we include a discussion on the situation when operating at cryogenic temperatures. 

The overlapping between optical and mechanical fields in Eq.(1) is deﬁned by the inner product between two vector ﬁelds with integration over the waveguide cross-section, S,

\begin{equation}
\langle \textbf{A},\textbf{B}\rangle=\int_{S}dS(\textbf{A}^{*}\cdot \textbf{B}) 
\end{equation}

In our case, the volume integration is reduced to surface integration over the transverse section of the waveguide because of the continuous translational symmetry along the propagation axis. We consider the total optical force $\textbf{f}$ exerted by pump and Stokes waves and originating from electrostriction bulk and interface forces as well as radiative pressure \cite{QIU13-OE}. The equations of elastodynamics, including the electrostrictive stress induced by the optical ﬁeld, reads \cite{LAU13-AIPA}:

\begin{equation}
\rho\frac{\partial^{2}u_{i}}{\partial t^{2}}-[c_{ijkl}u_{k,l}]_{j}=-\sigma^{es}_{ij,j}
\end{equation}

\noindent where $c_{ijkl}$ is the rank-4 tensor of elastic constants, $\sigma^{es}_{ij}=-\frac{1}{2}\epsilon_{0} \chi_{klij}E_{k}E_{l}$ is the electrostrictive stress tensor, with the rank-4 susceptibility tensor $\chi_{klij}=\epsilon_{km}\epsilon_{ln}p_{mnij}(\sigma_{ij}^{es}=-\frac{1}{2}\epsilon_{0}\epsilon _{km} \epsilon_{ln}p_{mnij}E_{k}E_{l})$ , $p_{mnij}$ is the photoelastic tensor, and $\epsilon_{0}$ is the permittivity of vacuum. 

Finally, the overlapping between optical and mechanical fields in Eq.(1) is defined as

\begin{eqnarray}
\langle \textbf{f},\textbf{u}\rangle=\iint_{S}dS(\textbf{f}^{ES,*}\cdot \textbf{u})+\oint_{\partial S}dl(\textbf{F}^{ES,*}\cdot \textbf{u}) \nonumber \\
+ \oint_{\partial S}dl(\textbf{F}^{RP,*}\cdot \textbf{u})
\end{eqnarray}

\noindent where $f^{ES}=-\nabla\widehat{\sigma}$ is the electrostriction (bulk) force, $\textbf{F}^{ES}=(\widehat{\sigma}^{(1)}-\widehat{\sigma}^{(2)})\cdot \textbf{n}$ is the electrostriction interface force and $\textbf{F}^{RP}=(\widehat{T}^{(1)}-\widehat{T}^{(2)})\cdot \textbf{n}$ is the radiative pressure, $\textbf{n}$ being the normal to the boundary between contacting media. Here the index (1) describes the internal medium and (2) describes the outward medium of the interface. The radiative pressure is defined by the Maxwell stress tensor, $T_{ij}=\varepsilon_{0}\varepsilon(E_{i}E_{j}-\frac{1}{2}\delta_{ij} E^{2} )$, and can be reduced to the expression:

\begin{equation}
\textbf{F}^{RP}=\frac{1}{2}(\varepsilon_{0}(\varepsilon_{1}-\varepsilon_{2})E_{pt}E_{st}^{*}-\varepsilon_{0}^{-1}(\varepsilon_{1}^{-1}-\varepsilon_{2}^{-1}))D_{pn}D_{sn}^{*})\cdot \textbf{n}
\end{equation}

\section{\label{sec:level1}Forward Stimulated Brillouin Scattering}

In the case of released silicon waveguides, mechanical modes are completely guided even when $q$ = 0 because phonons cannot escape from the silicon core. Therefore, large values of $G_{B}$ in Forward Stimulated Brillouin Scattering (FSBS) processes have been obtained by simulations and confirmed experimentally \cite{LAER15-NP}. When the silicon – or $Si_{3}N_{4}$ - core is on top of a silica substrate, there is no way to confine the phonons in the waveguide since the sound velocity in silica is lower than in silicon. In our system, even though mechanical modes at $q\approx$0 (required for FSBS) are leaky (or, in other words, they do not have an infinite mechanical Q factor ideally), we can get mechanical modes with a relatively large confinement of the displacement field inside the silicon core, as shown in Fig. 1(c). This is because the intermediate $Si_{3}N_{4}$ layer partially prevents the leakage of phonons into the silica substrate. Therefore, we have performed numerical calculations to evaluate $G_{B}$ for the FSBS process in two situations: intramodal TE and TM optical modes interaction.

\begin{figure}[htbp]
\includegraphics[width=0.47\textwidth]{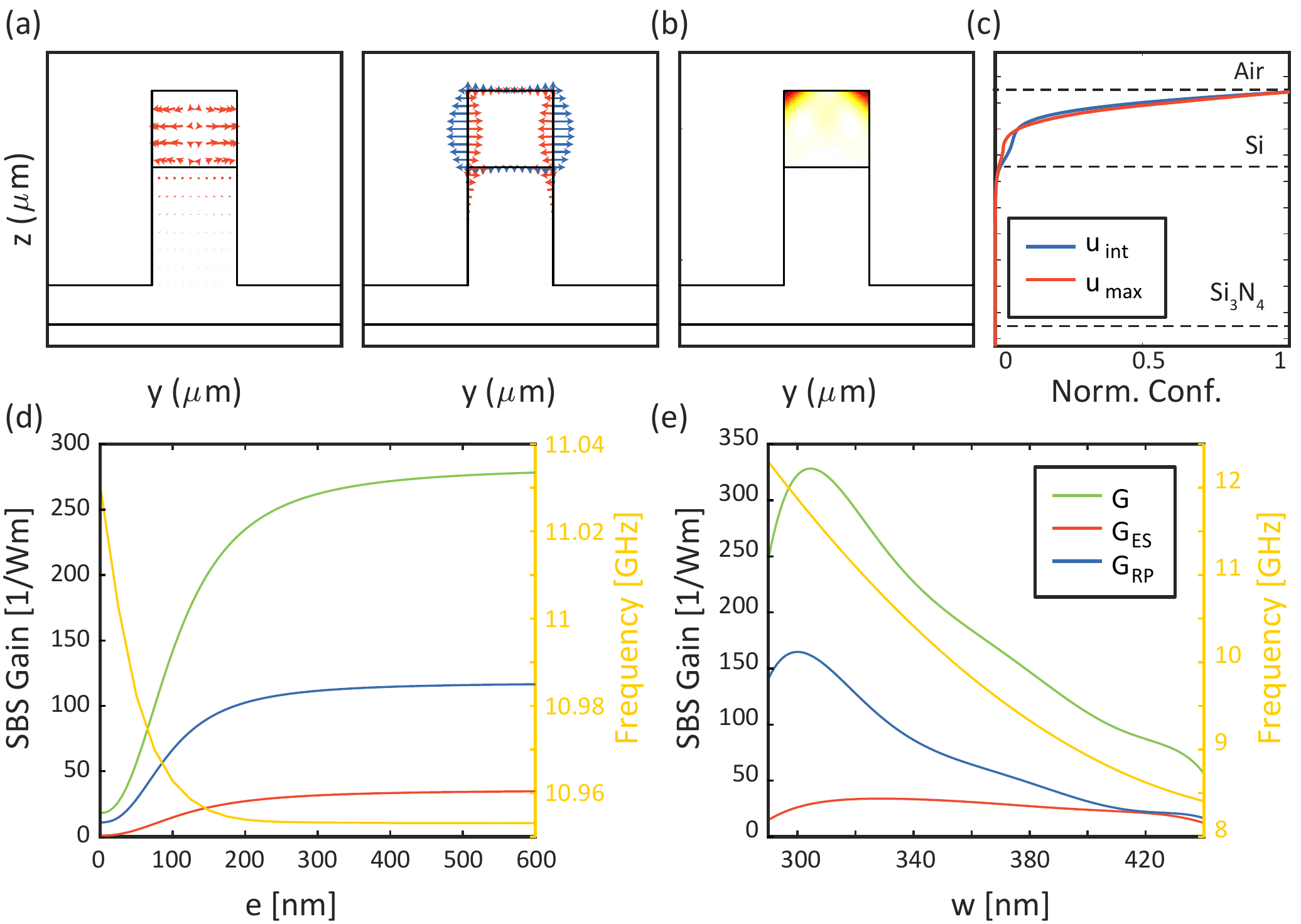}
\caption{Intramodal TE FSBS interaction at $q$=0. (a) Optical force distributions between the interaction of two TE-like modes for FSBS. The first panel is shown in red the bulk electrostrictive force and the second panel shows the electrostrictive (red) and radiation pressure (blue) interface forces. (b) Total displacement of the mode with the highest $G_{B}$ (termed A in Fig. \ref{fig:intro}(c)). (c) Normalized confinement (Norm. Conf.) distribution of the total accumulated displacement (blue) and maximum displacement (red) . (d) Total SBS Gain (green) and its electrostrictive (red) and radiation pressure (blue) contributions as a function of the etch $Si_{3}N_{4}$ layer for a waveguide width of $w$=325 nm, $b$=0.9$w$ and $t$=600 nm. (e) Total SBS Gain (green) and its electrostrictive (red) and radiation pressure (blue) contributions as a function of waveguide width for an etch height of $e$=450 nm and $t$=600 nm.}
\label{fig:fig2}
\end{figure}

The results for intramodal TE FSBS are summarized in Fig. \ref{fig:fig2}. Here, we can observe the obtained forces profiles in the case of the bulk and interface electrostrictive (red) and radiation (blue) pressure forces (Fig. \ref{fig:fig2}(a)). For these forces profiles, the mechanical mode which gives the highest SBS Gain is presented in Fig. \ref{fig:fig2}(b), where it can be appreciated that it is highly located at the top of the silica core. This can also be observed once we calculate both the total integrated or accumulated displacement (blue) and the maximum displacement (red) as a function of the height, as depicted on Fig. \ref{fig:fig2}(c). This mode has a total theoretical quality factor of $2\cdot 10^{5}$. The total accumulated displacement is calculated as $\int_{-w/2}^{w/2}\vert u\vert^{2}dy$. We can see from the trend of this figure that the mode is almost entirely confined in the $Si$ core and it has its maximum at the edges. In this case, we obtain relatively large $G_{B}$ values ($G_{B,TE}\approx$ 300 1/Wm) for the mechanical mode A as long as the etching e is larger than 250 nm (Fig. \ref{fig:fig2}(d)). Indeed, the general trend is that the SBS gain is higher for higher e values, which could be expected since a deeper etching would result in a better confined mechanical mode, and it saturates when $e$ increases. When calculating the Brillouin gain as a function of $w$, we can see that there is a waveguide width that maximizes the gain, as shown in \ref{fig:fig2}(e). 
Remarkably, changing the waveguide width also allows us to tune the mechanical frequency. As shown in Fig. \ref{fig:fig2}(e), changing the waveguide width from 290 to 450 nm tunes the mechanical frequency from $\approx$8 GHz to $\approx$12 GHz. This is an important feature because it may allow to tune the Brillouin frequency across the whole X band, which is very important in many microwave applications. As the mechanical mode is mainly confined on the edges of the upper surface of the silicon region, as can be appreciated in Fig. \ref{fig:fig2}(b), the main contribution to the total SBS gain is given by the radiation pressure force at that interface. Regarding the electrostrictive contribution it has to be noted that the field profiles of the bulk and interface contributions have opposite directions, resulting in a destructive interference of the total contribution. Nevertheless, both contributions, the radiation pressure and the total electrostrictive effect, are constructive. It must be noted that this mechanical mode has an average confinement of 0.989, as can be appreciated in Fig. \ref{fig:intro}(c) at $q$=0. This can explain the trend in the evolution of the SBS gain as a function of the waveguide width. As the width decreases, there is a higher overlap between this interface force and the mechanical field distribution, thus giving rise to higher SBS gains. 

\begin{figure}[htbp]
\includegraphics[scale=0.4]{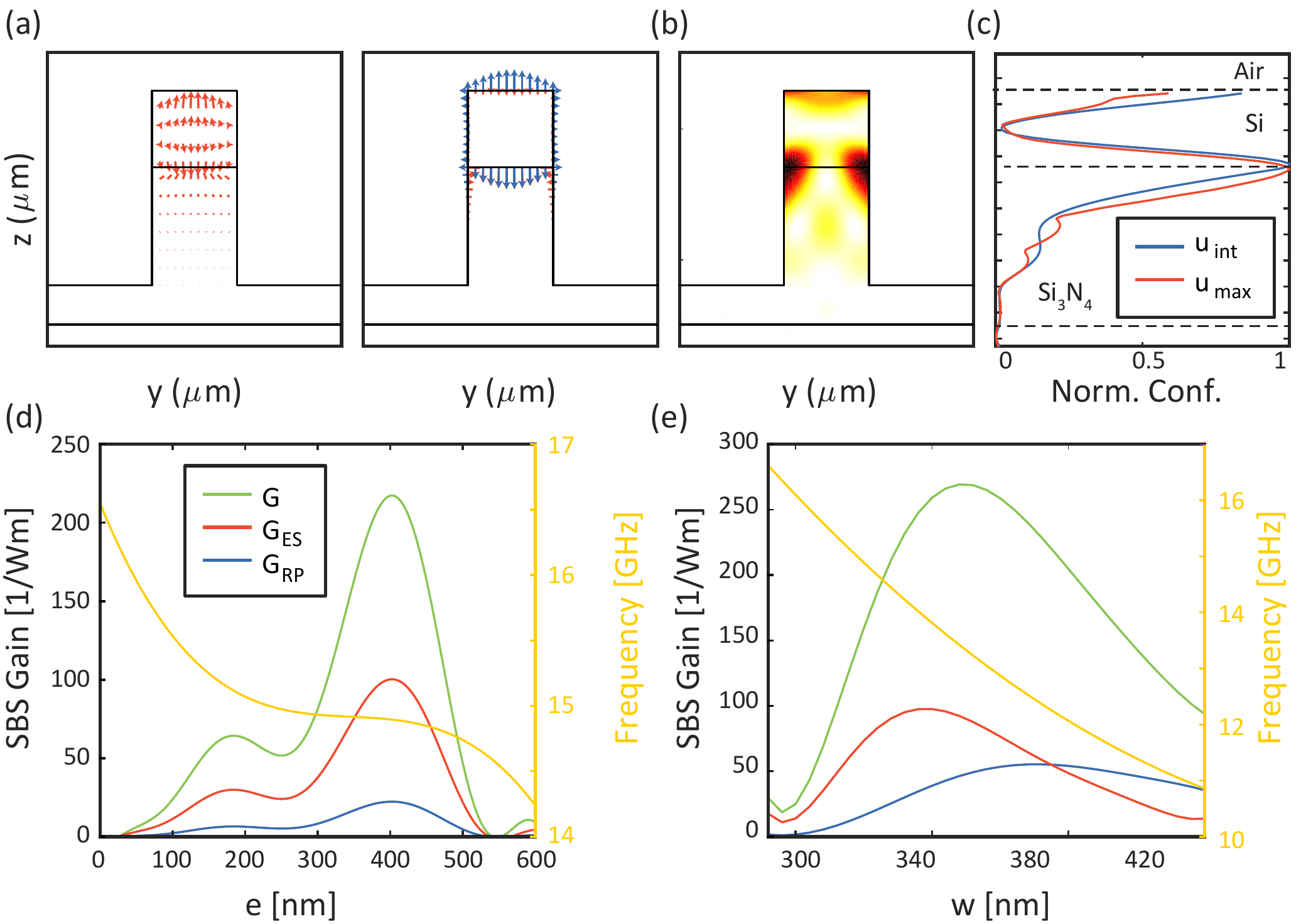}
\caption{Intramodal TM FSBS interaction at $q$=0. (a) Optical force distributions between the interaction of two TM-like modes for FSBS. The first panel shows in red the bulk electrostrictive force and the second panel shows the electrostrictive (red) and radiation pressure (blue) interface forces. (b) Total displacement of the mode with the highest $G_{B}$ (termed C in Fig. \ref{fig:intro}(c)). (c) Normalized confinement distribution of the total accumulated displacement (blue) and maximum displacement (red) . (d) Total SBS Gain (green) and its electrostrictive (red) and radiation pressure (blue) contributions as a function of the etch $Si_{3}N_{4}$ layer for a waveguide width of $w$=325 nm, $b$=0.9$w$ and $t$=600 nm. (e) Total SBS Gain (green) and its electrostrictive (red) and radiation pressure (blue) contributions as a function of waveguide width for an etch height of $e$=450 nm and $t$=600 nm.}
\label{fig:fig3}
\end{figure}

The results for intramodal TM FSBS are summarized in Fig. \ref{fig:fig3}. Again, we observe relatively large values of $G_{B}$ (Fig. \ref{fig:fig3}(d)), peaking  close to 300 1/Wm. As in the previous case, we also observe that changing the waveguide width allows us for tuning the mechanical frequency and optimizing the AO coupling. Indeed, the best configuration can be found for an $e\approx$ 400 nm and $w$ = 360 nm, with a mechanical frequency $f_{m}\approx$ 13 GHz. Here, the main contribution to the total SBS Gain comes from the electrostrictive effect where the bulk and interface field profiles have now the same direction in the region where the mode is localized. The main difference with respect to the TE case is that the mechanical mode that couples with the forces distributions (Fig. \ref{fig:fig3}(b)) in this configuration is not so well confined in the silicon region as before. This relatively poor confinement can also be observed in Fig. \ref{fig:fig3}(c) where we can see two main peaks of localization of the mechanical mode. Indeed, the average localization factor is $\approx$0.5, being part of the mechanical displacement lying in the pedestal region, and the theoretical mechanical quality factor is $10^{2}$. In this case, as the $Q_{m}$ is lower than $10^{3}$, the mechanical quality factor used in the calculation of the SBS Gain was the theoretical one. 

\begin{figure}[htbp]
\includegraphics[scale=0.4]{
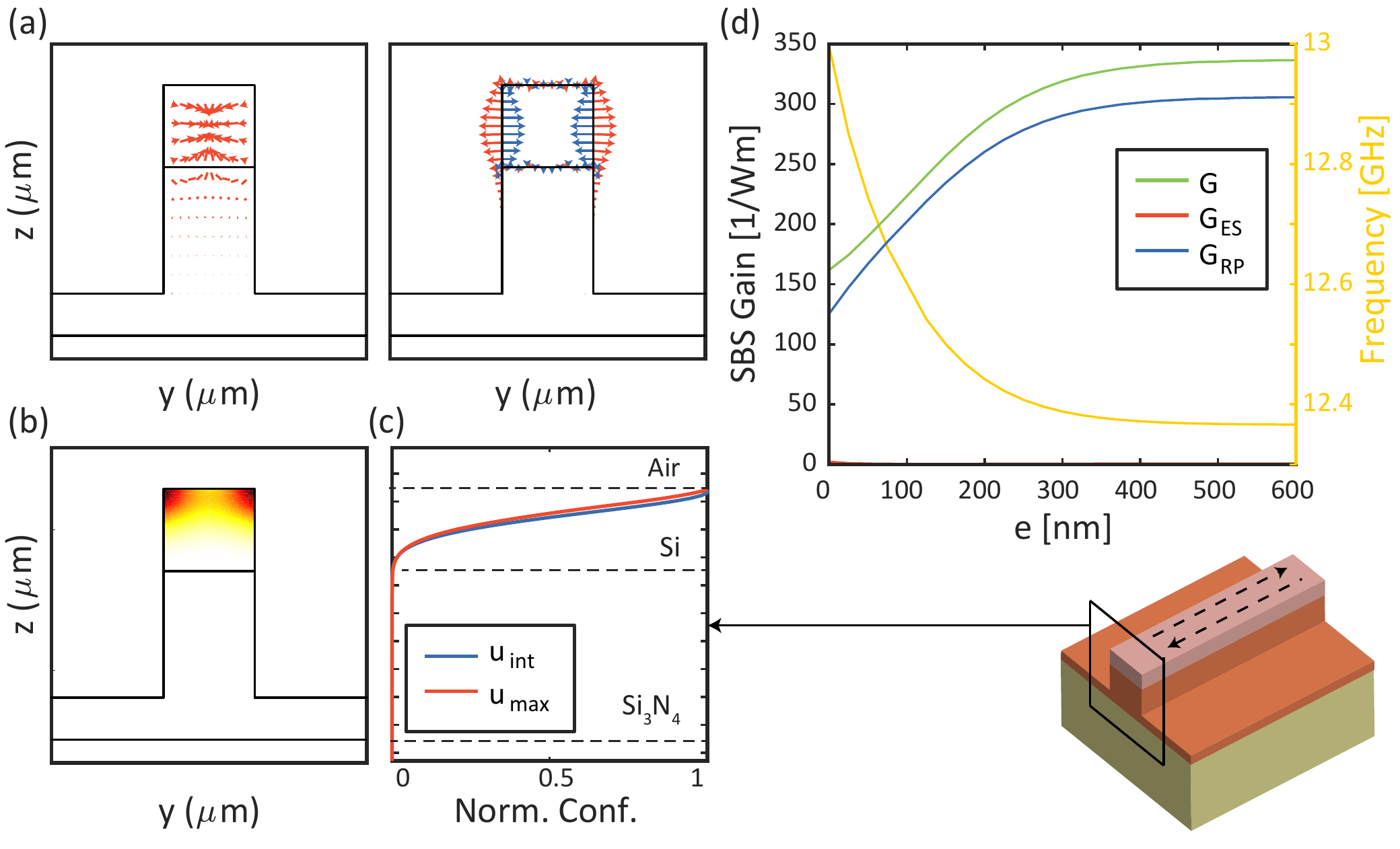}
\caption{Intramodal TE BSBS  interaction at $q$=$2k_{TE}$. (a) Optical force distributions between the interaction of two TE-like modes for BSBS. The first panel is shown in red the bulk electrostrictive force and the second panel shows the electrostrictive (red) and radiation pressure (blue) interface forces. (b) Total displacement of the mode with the highest $G_{B}$ (termed B in Fig. \ref{fig:intro}(c)). (c) Normalized confinement distribution of the total accumulated displacement (blue) and maximum displacement (red). (d) Total SBS Gain (green) and its electrostrictive (red) and radiation pressure (blue) contributions as a function of the etch $Si_{3}N_{4}$ layer for a waveguide width of $w$=325 nm, $b$=0.9$w$ and $t$=600 nm. }
\label{fig:fig4}
\end{figure}

\section{\label{sec:level1}Backward Stimulated Brillouin Scattering:}

The case of the Backward Stimulated Brillouin Scattering (BSBS) process involves large $q$ mechanical modes that should be ideally placed below the sound line of the silica substrate in order to prevent mechanical leakage. In our study, we consider first intramodal BSBS of the TE-like optical mode and select again the mechanical mode providing a larger $G_{B}$ (mode B in Fig. \ref{fig:intro}(c)). Again, we get values $G_{B}\approx$ 340  1/Wm for frequencies $f_{m}\approx$ 12.4 GHz for the dimensions specified in Fig. \ref{fig:fig4}(d). In this configuration, the electrostrictive contribution in the total SBS gain is nearly zero. This results from the destructive interference between both the bulk and the interface force that, as can be seen in Fig. \ref{fig:fig4}(b), are opposite. Besides that, we get a mechanical mode highly localized in the silicon core (Fig. \ref{fig:fig4}(b) and (c)). In this case, the theoretical mechanical quality factor is $2.8\cdot 10^{7}$, larger than in the forward configuration as expected.

\begin{figure}[htbp]
\includegraphics[scale=0.4]{
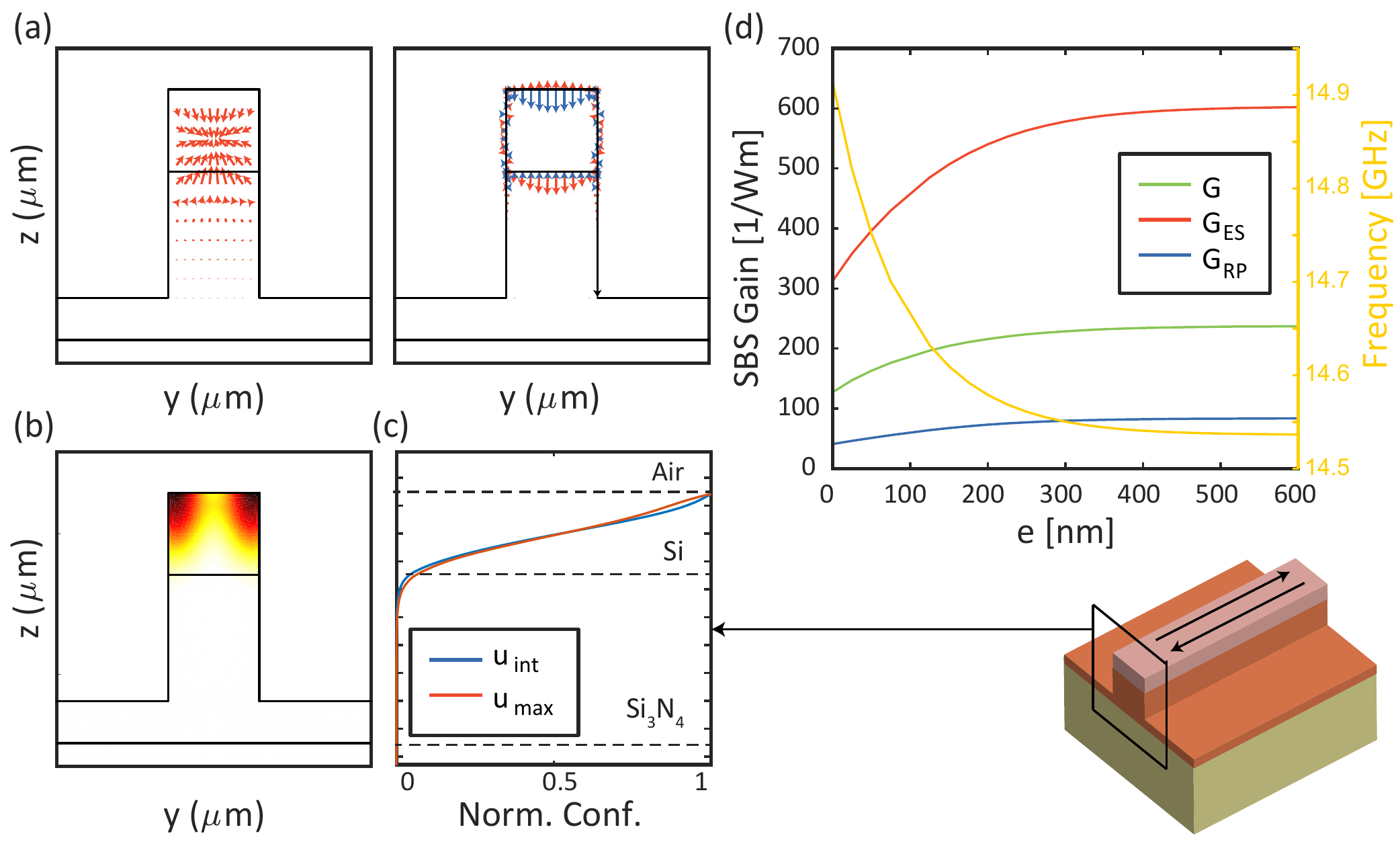}
\caption{Intramodal TM BSBS interaction at $q$=$2k_{TM}$. (a) Optical force distributions between the interaction of two TM-like modes for BSBS. The first panel is shown in red the bulk electrostrictive force and the second panel shows the electrostrictive (red) and radiation pressure (blue) interface forces. (b) Total displacement of the mode with the highest GB (termed B in Fig. \ref{fig:intro}(c)). (e) Normalized confinement distribution of the total accumulated displacement (blue) and maximum displacement (red). (d) SBS Gain as a function of the etch $Si_{3}N_{4}$ layer for a waveguide width of $w$=325 nm, $b$=0.9$w$ and $t$=600 nm. (e) Total SBS Gain (green) and its electrostrictive (red) and radiation pressure (blue) contributions as a function of waveguide width layer for a an etch height of $e$=450 nm and $t$=600 nm.}
\label{fig:fig5}
\end{figure}

In the case of backward intramodal TM  configuration (Fig. \ref{fig:fig5}), we observe that unlike in intramodal TE BSBS case, now the ES contribution gain reaches values of 600 1/Wm.  However, the RP and ES contributions have a destructive interference, thus resulting in $G_{B}\approx$ 220 1/Wm, also for the mode termed B in Fig. \ref{fig:intro}(c). Nevertheless, as the mechanical mode has a high localization factor, we observe a slight improvement in the SBS gain when the etching depth increases. Still, these values could be sufficient for practical applications. The mechanical quality factor of the mode depicted in Fig. \ref{fig:fig5}(b) is $2.3\cdot 10^{7}$.

Finally, regarding the intermodal TE-TM interaction the values of the SBS gain were lower than in the case of intramodal interactions. This is probably due to the fact that the mode that gives the highest SBS Gain for this configuration at $q=k_{TM}-k_{TE}$ is the mode termed D in Fig. \ref{fig:intro}(c), which has a low localization even at high wavevectors.

\section{\label{sec:level1}Low temperature SBS Gain enhancement:}

As we noted in previous sections, the mechanical Q factor was assumed to be $Q_{m}=10^{3}$ because of the mechanical damping is limited by the material loss at room temperature \cite{MAC19-ARX}. However, this is not the case if we work at low temperatures, where the mechanical quality factor can be enhanced up to much higher values. 

\begin{figure}[htbp]
\includegraphics[width=0.47\textwidth]{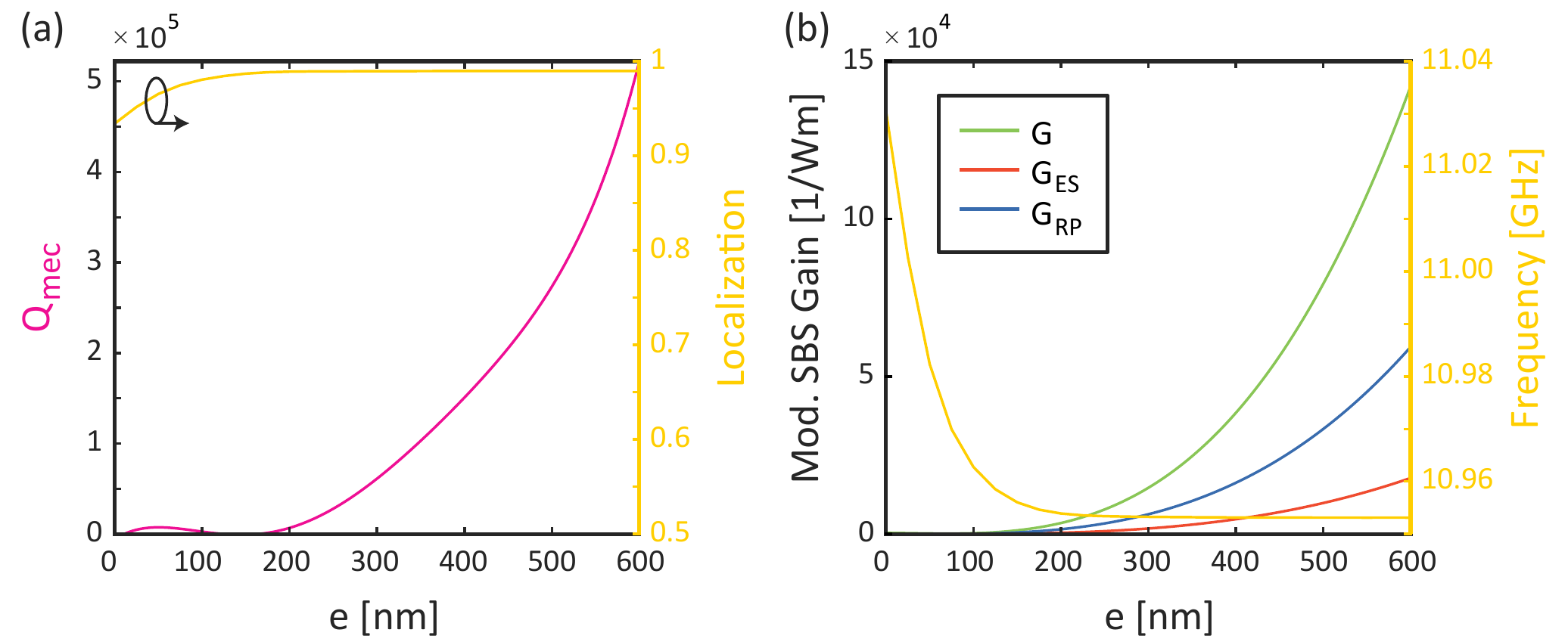}
\caption{Intramodal TE FSBS  interaction at $q$=0 with the theoretical quality factor . (a) Mechanical quality factor (purple) and localization in the silicon core (yellow) as a function of the etch $Si_{3}N_{4}$ layer. (b) SBS Gain as a function of the etch $Si_{3}N_{4}$ layer for a waveguide width of $w$=325 nm, $b$=0.9$w$ and $t$=600 nm with the theoretical quality factor in (a). }
\label{fig:fig6}
\end{figure}

If we study the evolution of the ideal mechanical quality factor (this is, as obtained from the simulations and neglecting material losses) of the mechanical modes involved in our configurations, we can see that this factor can be enhanced as the etching of the $Si_{3}N_{4}$ layer is increased, with constant  $Si_{3}N_{4}$ thickness as t=600 nm, as can be seen in Fig. \ref{fig:fig6}(a) for the case of the intramodal TE FSBS  interaction. Here, we can see also the evolution of the localization factor (in yellow) in the silicon core. It can be appreciated that the mechanical mode is almost totally confined (the localization factor is almost 1) for all the etch ratios and it is only reduced in the case where there is not any pedestal height and the silicon waveguide lies into the $Si_{3}N_{4}$ substrate. The modified SBS Gain, taking into account this theoretical quality factor and introducing it in Eq. \ref{eq:sbs1} provides the values in Fig. \ref{fig:fig6}(b). This means that for the same case as in Fig. \ref{fig:fig2}(d) we have been able to improve the total SBS gain from 300 1/Wm up to 1.5$\cdot 10^{5}$ 1/Wm, which should be attainable when operating at cryogenic temperatures.

\section{\label{sec:level1}Conclusions:}

In summary, we have shown a novel way towards unreleased cavity and waveguide optomechanics in silicon technology. The presented structure relies on vertical engineering: a thick layer of silicon nitride separated the silicon core from the silica substrate, reducing the mechanical leakage whilst enabling photonic guidance in the silicon core. Values of the Brillouin gain around 300 1/Wm can be achieved for several configurations and mechanical modes above 10 GHz when a mechanical Q factor equal to $10^{3}$ is considered, which is usually the case in room-temperature optomechanics in silicon. Much higher values of the Brillouin gain are attainable if we consider operation at low temperature because of the decrease of the material losses. Optomechanical structures based on this multilayer system could be fabricated using standard silicon technology. Our finding could pave the way towards unreleased optomechanical circtuis, which would overcome the limitations arising from the need to release the silicon core in standard optomechanical circuitry. 

\begin{acknowledgments}
This work was supported by the European Commission (PHENOMEN H2020-EU-713450); Programa de Ayudas de Investigación y Desarrolo (PAID-01-16) de la Universitat Politècnica de València; Ministerio
de Ciencia, Innovación y Universidades (PGC2018- 094490-B, PRX18/00126) and Generalitat Valenciana (PROMETEO/2019/123). The authors thank Jouni Ahopelto for helpful comments.
\end{acknowledgments}

\bibliography{biblio}

\begin{thebibliography}{30}%
\makeatletter
\providecommand \@ifxundefined [1]{%
 \@ifx{#1\undefined}
}%
\providecommand \@ifnum [1]{%
 \ifnum #1\expandafter \@firstoftwo
 \else \expandafter \@secondoftwo
 \fi
}%
\providecommand \@ifx [1]{%
 \ifx #1\expandafter \@firstoftwo
 \else \expandafter \@secondoftwo
 \fi
}%
\providecommand \natexlab [1]{#1}%
\providecommand \enquote  [1]{``#1''}%
\providecommand \bibnamefont  [1]{#1}%
\providecommand \bibfnamefont [1]{#1}%
\providecommand \citenamefont [1]{#1}%
\providecommand \href@noop [0]{\@secondoftwo}%
\providecommand \href [0]{\begingroup \@sanitize@url \@href}%
\providecommand \@href[1]{\@@startlink{#1}\@@href}%
\providecommand \@@href[1]{\endgroup#1\@@endlink}%
\providecommand \@sanitize@url [0]{\catcode `\\12\catcode `\$12\catcode
  `\&12\catcode `\#12\catcode `\^12\catcode `\_12\catcode `\%12\relax}%
\providecommand \@@startlink[1]{}%
\providecommand \@@endlink[0]{}%
\providecommand \url  [0]{\begingroup\@sanitize@url \@url }%
\providecommand \@url [1]{\endgroup\@href {#1}{\urlprefix }}%
\providecommand \urlprefix  [0]{URL }%
\providecommand \Eprint [0]{\href }%
\providecommand \doibase [0]{https://doi.org/}%
\providecommand \selectlanguage [0]{\@gobble}%
\providecommand \bibinfo  [0]{\@secondoftwo}%
\providecommand \bibfield  [0]{\@secondoftwo}%
\providecommand \translation [1]{[#1]}%
\providecommand \BibitemOpen [0]{}%
\providecommand \bibitemStop [0]{}%
\providecommand \bibitemNoStop [0]{.\EOS\space}%
\providecommand \EOS [0]{\spacefactor3000\relax}%
\providecommand \BibitemShut  [1]{\csname bibitem#1\endcsname}%
\let\auto@bib@innerbib\@empty
\bibitem [{\citenamefont {{B}rillouin}(1922)}]{BRI22-AP}%
  \BibitemOpen
  \bibfield  {author} {\bibinfo {author} {\bibfnamefont {L.}~\bibnamefont
  {{B}rillouin}},\ }\bibfield  {title} {\bibinfo {title} {Diffusion de la
  lumière et des rayons x par un corps transparent homogène},\ }\href
  {https://doi.org/https://doi.org/10.1051/anphys/192209170088} {\bibfield
  {journal} {\bibinfo  {journal} {Ann. Phys.}\ }\textbf {\bibinfo {volume}
  {9}},\ \bibinfo {pages} {88 } (\bibinfo {year} {1922})}\BibitemShut {NoStop}%
\bibitem [{\citenamefont {Chiao}\ \emph {et~al.}(1964)\citenamefont {Chiao},
  \citenamefont {Townes},\ and\ \citenamefont {Stoicheff}}]{CHI64-PRL}%
  \BibitemOpen
  \bibfield  {author} {\bibinfo {author} {\bibfnamefont {R.~Y.}\ \bibnamefont
  {Chiao}}, \bibinfo {author} {\bibfnamefont {C.~H.}\ \bibnamefont {Townes}},\
  and\ \bibinfo {author} {\bibfnamefont {B.~P.}\ \bibnamefont {Stoicheff}},\
  }\bibfield  {title} {\bibinfo {title} {Stimulated {B}rillouin scattering and
  coherent generation of intense hypersonic waves},\ }\href
  {https://doi.org/10.1103/PhysRevLett.12.592} {\bibfield  {journal} {\bibinfo
  {journal} {Phys. Rev. Lett.}\ }\textbf {\bibinfo {volume} {12}},\ \bibinfo
  {pages} {592} (\bibinfo {year} {1964})}\BibitemShut {NoStop}%
\bibitem [{\citenamefont {Ippen}\ and\ \citenamefont
  {Stolen}(1972)}]{IPP72-APL}%
  \BibitemOpen
  \bibfield  {author} {\bibinfo {author} {\bibfnamefont {E.~P.}\ \bibnamefont
  {Ippen}}\ and\ \bibinfo {author} {\bibfnamefont {R.~H.}\ \bibnamefont
  {Stolen}},\ }\bibfield  {title} {\bibinfo {title} {Stimulated {B}rillouin
  scattering in optical fibers},\ }\href {https://doi.org/10.1063/1.1654249}
  {\bibfield  {journal} {\bibinfo  {journal} {Appl. Phys. Lett.}\ }\textbf
  {\bibinfo {volume} {21}},\ \bibinfo {pages} {539} (\bibinfo {year}
  {1972})}\BibitemShut {NoStop}%
\bibitem [{\citenamefont {Capmany}\ and\ \citenamefont
  {Novak}(2007)}]{CAP07-NP}%
  \BibitemOpen
  \bibfield  {author} {\bibinfo {author} {\bibfnamefont {J.}~\bibnamefont
  {Capmany}}\ and\ \bibinfo {author} {\bibfnamefont {D.}~\bibnamefont
  {Novak}},\ }\bibfield  {title} {\bibinfo {title} {Microwave photonics
  combines two worlds},\ }\href {https://doi.org/10.1038/nphoton.2007.89}
  {\bibfield  {journal} {\bibinfo  {journal} {Nature Photonics}\ }\textbf
  {\bibinfo {volume} {1}},\ \bibinfo {pages} {319} (\bibinfo {year}
  {2007})}\BibitemShut {NoStop}%
\bibitem [{\citenamefont {Vidal}\ \emph {et~al.}(2007)\citenamefont {Vidal},
  \citenamefont {Piqueras},\ and\ \citenamefont {Mart\'{i}}}]{VI07-OL}%
  \BibitemOpen
  \bibfield  {author} {\bibinfo {author} {\bibfnamefont {B.}~\bibnamefont
  {Vidal}}, \bibinfo {author} {\bibfnamefont {M.~A.}\ \bibnamefont
  {Piqueras}},\ and\ \bibinfo {author} {\bibfnamefont {J.}~\bibnamefont
  {Mart\'{i}}},\ }\bibfield  {title} {\bibinfo {title} {Tunable and
  reconfigurable photonic microwave filter based on stimulated {B}rillouin
  scattering},\ }\href {https://doi.org/10.1364/OL.32.000023} {\bibfield
  {journal} {\bibinfo  {journal} {Opt. Lett.}\ }\textbf {\bibinfo {volume}
  {32}},\ \bibinfo {pages} {23} (\bibinfo {year} {2007})}\BibitemShut {NoStop}%
\bibitem [{\citenamefont {Zhu}\ \emph {et~al.}(2007)\citenamefont {Zhu},
  \citenamefont {Gauthier},\ and\ \citenamefont {Boyd}}]{ZHU07-SCI}%
  \BibitemOpen
  \bibfield  {author} {\bibinfo {author} {\bibfnamefont {Z.}~\bibnamefont
  {Zhu}}, \bibinfo {author} {\bibfnamefont {D.~J.}\ \bibnamefont {Gauthier}},\
  and\ \bibinfo {author} {\bibfnamefont {R.~W.}\ \bibnamefont {Boyd}},\
  }\bibfield  {title} {\bibinfo {title} {Stored light in an optical fiber via
  stimulated {B}rillouin scattering},\ }\href
  {https://doi.org/10.1126/science.1149066} {\bibfield  {journal} {\bibinfo
  {journal} {Science}\ }\textbf {\bibinfo {volume} {318}},\ \bibinfo {pages}
  {1748} (\bibinfo {year} {2007})}\BibitemShut {NoStop}%
\bibitem [{\citenamefont {Eggleton}\ \emph {et~al.}(2019)\citenamefont
  {Eggleton}, \citenamefont {Poulton}, \citenamefont {Rakich}, \citenamefont
  {Steel},\ and\ \citenamefont {Bahl}}]{EGG19-NP}%
  \BibitemOpen
  \bibfield  {author} {\bibinfo {author} {\bibfnamefont {B.~J.}\ \bibnamefont
  {Eggleton}}, \bibinfo {author} {\bibfnamefont {C.~G.}\ \bibnamefont
  {Poulton}}, \bibinfo {author} {\bibfnamefont {P.~T.}\ \bibnamefont {Rakich}},
  \bibinfo {author} {\bibfnamefont {M.~J.}\ \bibnamefont {Steel}},\ and\
  \bibinfo {author} {\bibfnamefont {G.}~\bibnamefont {Bahl}},\ }\bibfield
  {title} {\bibinfo {title} {{B}rillouin integrated photonics},\ }\href
  {https://doi.org/10.1038/s41566-019-0498-z} {\bibfield  {journal} {\bibinfo
  {journal} {Nature Photonics}\ }\textbf {\bibinfo {volume} {13}},\ \bibinfo
  {pages} {664} (\bibinfo {year} {2019})}\BibitemShut {NoStop}%
\bibitem [{\citenamefont {Pant}\ \emph {et~al.}(2011)\citenamefont {Pant},
  \citenamefont {Poulton}, \citenamefont {Choi}, \citenamefont {Mcfarlane},
  \citenamefont {Hile}, \citenamefont {Li}, \citenamefont {Thevenaz},
  \citenamefont {Luther-Davies}, \citenamefont {Madden},\ and\ \citenamefont
  {Eggleton}}]{PANT11-OE}%
  \BibitemOpen
  \bibfield  {author} {\bibinfo {author} {\bibfnamefont {R.}~\bibnamefont
  {Pant}}, \bibinfo {author} {\bibfnamefont {C.~G.}\ \bibnamefont {Poulton}},
  \bibinfo {author} {\bibfnamefont {D.-Y.}\ \bibnamefont {Choi}}, \bibinfo
  {author} {\bibfnamefont {H.}~\bibnamefont {Mcfarlane}}, \bibinfo {author}
  {\bibfnamefont {S.}~\bibnamefont {Hile}}, \bibinfo {author} {\bibfnamefont
  {E.}~\bibnamefont {Li}}, \bibinfo {author} {\bibfnamefont {L.}~\bibnamefont
  {Thevenaz}}, \bibinfo {author} {\bibfnamefont {B.}~\bibnamefont
  {Luther-Davies}}, \bibinfo {author} {\bibfnamefont {S.~J.}\ \bibnamefont
  {Madden}},\ and\ \bibinfo {author} {\bibfnamefont {B.~J.}\ \bibnamefont
  {Eggleton}},\ }\bibfield  {title} {\bibinfo {title} {On-chip stimulated
  {B}rillouin scattering},\ }\href {https://doi.org/10.1364/OE.19.008285}
  {\bibfield  {journal} {\bibinfo  {journal} {Opt. Express}\ }\textbf {\bibinfo
  {volume} {19}},\ \bibinfo {pages} {8285} (\bibinfo {year}
  {2011})}\BibitemShut {NoStop}%
\bibitem [{\citenamefont {Pennec}\ \emph {et~al.}(2014)\citenamefont {Pennec},
  \citenamefont {Laude}, \citenamefont {Papanikolaou}, \citenamefont
  {Djafari-Rouhani}, \citenamefont {Oudich}, \citenamefont {Jallal},
  \citenamefont {Beugnot}, \citenamefont {Escalante},\ and\ \citenamefont
  {Martínez}}]{PEN14-NP}%
  \BibitemOpen
  \bibfield  {author} {\bibinfo {author} {\bibfnamefont {Y.}~\bibnamefont
  {Pennec}}, \bibinfo {author} {\bibfnamefont {V.}~\bibnamefont {Laude}},
  \bibinfo {author} {\bibfnamefont {N.}~\bibnamefont {Papanikolaou}}, \bibinfo
  {author} {\bibfnamefont {B.}~\bibnamefont {Djafari-Rouhani}}, \bibinfo
  {author} {\bibfnamefont {M.}~\bibnamefont {Oudich}}, \bibinfo {author}
  {\bibfnamefont {S.~E.}\ \bibnamefont {Jallal}}, \bibinfo {author}
  {\bibfnamefont {J.~C.}\ \bibnamefont {Beugnot}}, \bibinfo {author}
  {\bibfnamefont {J.~M.}\ \bibnamefont {Escalante}},\ and\ \bibinfo {author}
  {\bibfnamefont {A.}~\bibnamefont {Martínez}},\ }\bibfield  {title} {\bibinfo
  {title} {Modeling light-sound interaction in nanoscale cavities and
  waveguides},\ }\href
  {https://doi.org/https://doi.org/10.1515/nanoph-2014-0004} {\bibfield
  {journal} {\bibinfo  {journal} {Nanophotonics}\ }\textbf {\bibinfo {volume}
  {3}},\ \bibinfo {pages} {413 } (\bibinfo {year} {2014})}\BibitemShut
  {NoStop}%
\bibitem [{\citenamefont {{Choudhary}}\ \emph {et~al.}(2017)\citenamefont
  {{Choudhary}}, \citenamefont {{Morrison}}, \citenamefont {{Aryanfar}},
  \citenamefont {{Shahnia}}, \citenamefont {{Pagani}}, \citenamefont {{Liu}},
  \citenamefont {{Vu}}, \citenamefont {{Madden}}, \citenamefont {{Marpaung}},\
  and\ \citenamefont {{Eggleton}}}]{CHO17-JLT}%
  \BibitemOpen
  \bibfield  {author} {\bibinfo {author} {\bibfnamefont {A.}~\bibnamefont
  {{Choudhary}}}, \bibinfo {author} {\bibfnamefont {B.}~\bibnamefont
  {{Morrison}}}, \bibinfo {author} {\bibfnamefont {I.}~\bibnamefont
  {{Aryanfar}}}, \bibinfo {author} {\bibfnamefont {S.}~\bibnamefont
  {{Shahnia}}}, \bibinfo {author} {\bibfnamefont {M.}~\bibnamefont {{Pagani}}},
  \bibinfo {author} {\bibfnamefont {Y.}~\bibnamefont {{Liu}}}, \bibinfo
  {author} {\bibfnamefont {K.}~\bibnamefont {{Vu}}}, \bibinfo {author}
  {\bibfnamefont {S.}~\bibnamefont {{Madden}}}, \bibinfo {author}
  {\bibfnamefont {D.}~\bibnamefont {{Marpaung}}},\ and\ \bibinfo {author}
  {\bibfnamefont {B.~J.}\ \bibnamefont {{Eggleton}}},\ }\bibfield  {title}
  {\bibinfo {title} {Advanced integrated microwave signal processing with giant
  on-chip {B}rillouin gain},\ }\href@noop {} {\bibfield  {journal} {\bibinfo
  {journal} {Journal of Lightwave Technology}\ }\textbf {\bibinfo {volume}
  {35}},\ \bibinfo {pages} {846} (\bibinfo {year} {2017})}\BibitemShut
  {NoStop}%
\bibitem [{\citenamefont {Marpaung}\ \emph {et~al.}(2015)\citenamefont
  {Marpaung}, \citenamefont {Morrison}, \citenamefont {Pagani}, \citenamefont
  {Pant}, \citenamefont {Choi}, \citenamefont {Luther-Davies}, \citenamefont
  {Madden},\ and\ \citenamefont {Eggleton}}]{MAR15-OPT}%
  \BibitemOpen
  \bibfield  {author} {\bibinfo {author} {\bibfnamefont {D.}~\bibnamefont
  {Marpaung}}, \bibinfo {author} {\bibfnamefont {B.}~\bibnamefont {Morrison}},
  \bibinfo {author} {\bibfnamefont {M.}~\bibnamefont {Pagani}}, \bibinfo
  {author} {\bibfnamefont {R.}~\bibnamefont {Pant}}, \bibinfo {author}
  {\bibfnamefont {D.-Y.}\ \bibnamefont {Choi}}, \bibinfo {author}
  {\bibfnamefont {B.}~\bibnamefont {Luther-Davies}}, \bibinfo {author}
  {\bibfnamefont {S.~J.}\ \bibnamefont {Madden}},\ and\ \bibinfo {author}
  {\bibfnamefont {B.~J.}\ \bibnamefont {Eggleton}},\ }\bibfield  {title}
  {\bibinfo {title} {Low-power, chip-based stimulated {B}rillouin scattering
  microwave photonic filter with ultrahigh selectivity},\ }\href
  {https://doi.org/10.1364/OPTICA.2.000076} {\bibfield  {journal} {\bibinfo
  {journal} {Optica}\ }\textbf {\bibinfo {volume} {2}},\ \bibinfo {pages} {76}
  (\bibinfo {year} {2015})}\BibitemShut {NoStop}%
\bibitem [{\citenamefont {Safavi-Naeini}\ \emph {et~al.}(2019)\citenamefont
  {Safavi-Naeini}, \citenamefont {Thourhout}, \citenamefont {Baets},\ and\
  \citenamefont {Laer}}]{SAF19-OPT}%
  \BibitemOpen
  \bibfield  {author} {\bibinfo {author} {\bibfnamefont {A.~H.}\ \bibnamefont
  {Safavi-Naeini}}, \bibinfo {author} {\bibfnamefont {D.~V.}\ \bibnamefont
  {Thourhout}}, \bibinfo {author} {\bibfnamefont {R.}~\bibnamefont {Baets}},\
  and\ \bibinfo {author} {\bibfnamefont {R.~V.}\ \bibnamefont {Laer}},\
  }\bibfield  {title} {\bibinfo {title} {Controlling phonons and photons at the
  wavelength scale: integrated photonics meets integrated phononics},\ }\href
  {https://doi.org/10.1364/OPTICA.6.000213} {\bibfield  {journal} {\bibinfo
  {journal} {Optica}\ }\textbf {\bibinfo {volume} {6}},\ \bibinfo {pages} {213}
  (\bibinfo {year} {2019})}\BibitemShut {NoStop}%
\bibitem [{\citenamefont {Shin}\ \emph {et~al.}(2013)\citenamefont {Shin},
  \citenamefont {Qiu}, \citenamefont {Jarecki}, \citenamefont {Cox},
  \citenamefont {Olsson}, \citenamefont {Starbuck}, \citenamefont {Wang},\ and\
  \citenamefont {Rakich}}]{SHIN13-NCOMM}%
  \BibitemOpen
  \bibfield  {author} {\bibinfo {author} {\bibfnamefont {H.}~\bibnamefont
  {Shin}}, \bibinfo {author} {\bibfnamefont {W.}~\bibnamefont {Qiu}}, \bibinfo
  {author} {\bibfnamefont {R.}~\bibnamefont {Jarecki}}, \bibinfo {author}
  {\bibfnamefont {J.~A.}\ \bibnamefont {Cox}}, \bibinfo {author} {\bibfnamefont
  {R.~H.}\ \bibnamefont {Olsson}}, \bibinfo {author} {\bibfnamefont
  {A.}~\bibnamefont {Starbuck}}, \bibinfo {author} {\bibfnamefont
  {Z.}~\bibnamefont {Wang}},\ and\ \bibinfo {author} {\bibfnamefont {P.~T.}\
  \bibnamefont {Rakich}},\ }\bibfield  {title} {\bibinfo {title} {Tailorable
  stimulated {B}rillouin scattering in nanoscale silicon waveguides},\ }\href
  {https://doi.org/10.1038/ncomms2943} {\bibfield  {journal} {\bibinfo
  {journal} {Nature Communications}\ }\textbf {\bibinfo {volume} {4}},\
  \bibinfo {pages} {1944} (\bibinfo {year} {2013})}\BibitemShut {NoStop}%
\bibitem [{\citenamefont {Van~Laer}\ \emph {et~al.}(2015)\citenamefont
  {Van~Laer}, \citenamefont {Kuyken}, \citenamefont {Van~Thourhout},\ and\
  \citenamefont {Baets}}]{LAER15-NP}%
  \BibitemOpen
  \bibfield  {author} {\bibinfo {author} {\bibfnamefont {R.}~\bibnamefont
  {Van~Laer}}, \bibinfo {author} {\bibfnamefont {B.}~\bibnamefont {Kuyken}},
  \bibinfo {author} {\bibfnamefont {D.}~\bibnamefont {Van~Thourhout}},\ and\
  \bibinfo {author} {\bibfnamefont {R.}~\bibnamefont {Baets}},\ }\bibfield
  {title} {\bibinfo {title} {Interaction between light and highly confined
  hypersound in a silicon photonic nanowire},\ }\href
  {https://doi.org/10.1038/nphoton.2015.11} {\bibfield  {journal} {\bibinfo
  {journal} {Nature Photonics}\ }\textbf {\bibinfo {volume} {9}},\ \bibinfo
  {pages} {199} (\bibinfo {year} {2015})}\BibitemShut {NoStop}%
\bibitem [{\citenamefont {Laer}\ \emph {et~al.}(2015)\citenamefont {Laer},
  \citenamefont {Bazin}, \citenamefont {Kuyken}, \citenamefont {Baets},\ and\
  \citenamefont {Thourhout}}]{LAER15-NJP}%
  \BibitemOpen
  \bibfield  {author} {\bibinfo {author} {\bibfnamefont {R.~V.}\ \bibnamefont
  {Laer}}, \bibinfo {author} {\bibfnamefont {A.}~\bibnamefont {Bazin}},
  \bibinfo {author} {\bibfnamefont {B.}~\bibnamefont {Kuyken}}, \bibinfo
  {author} {\bibfnamefont {R.}~\bibnamefont {Baets}},\ and\ \bibinfo {author}
  {\bibfnamefont {D.~V.}\ \bibnamefont {Thourhout}},\ }\bibfield  {title}
  {\bibinfo {title} {Net on-chip {B}rillouin gain based on suspended silicon
  nanowires},\ }\href {http://dx.doi.org/10.1088/1367-2630/17/11/115005}
  {\bibfield  {journal} {\bibinfo  {journal} {New Journal of Physics}\ }\textbf
  {\bibinfo {volume} {17}},\ \bibinfo {pages} {115005} (\bibinfo {year}
  {2015})}\BibitemShut {NoStop}%
\bibitem [{\citenamefont {Chan}\ \emph {et~al.}(2012)\citenamefont {Chan},
  \citenamefont {Safavi-Naeini}, \citenamefont {Hill}, \citenamefont
  {Meenehan},\ and\ \citenamefont {Painter}}]{CHAN12-APL}%
  \BibitemOpen
  \bibfield  {author} {\bibinfo {author} {\bibfnamefont {J.}~\bibnamefont
  {Chan}}, \bibinfo {author} {\bibfnamefont {A.~H.}\ \bibnamefont
  {Safavi-Naeini}}, \bibinfo {author} {\bibfnamefont {J.~T.}\ \bibnamefont
  {Hill}}, \bibinfo {author} {\bibfnamefont {S.}~\bibnamefont {Meenehan}},\
  and\ \bibinfo {author} {\bibfnamefont {O.}~\bibnamefont {Painter}},\
  }\bibfield  {title} {\bibinfo {title} {Optimized optomechanical crystal
  cavity with acoustic radiation shield},\ }\href
  {https://doi.org/10.1063/1.4747726} {\bibfield  {journal} {\bibinfo
  {journal} {Applied Physics Letters}\ }\textbf {\bibinfo {volume} {101}},\
  \bibinfo {pages} {081115} (\bibinfo {year} {2012})}\BibitemShut {NoStop}%
\bibitem [{\citenamefont {Otterstrom}\ \emph {et~al.}(2018)\citenamefont
  {Otterstrom}, \citenamefont {Behunin}, \citenamefont {Kittlaus},
  \citenamefont {Wang},\ and\ \citenamefont {Rakich}}]{OTT18-SCI}%
  \BibitemOpen
  \bibfield  {author} {\bibinfo {author} {\bibfnamefont {N.~T.}\ \bibnamefont
  {Otterstrom}}, \bibinfo {author} {\bibfnamefont {R.~O.}\ \bibnamefont
  {Behunin}}, \bibinfo {author} {\bibfnamefont {E.~A.}\ \bibnamefont
  {Kittlaus}}, \bibinfo {author} {\bibfnamefont {Z.}~\bibnamefont {Wang}},\
  and\ \bibinfo {author} {\bibfnamefont {P.~T.}\ \bibnamefont {Rakich}},\
  }\bibfield  {title} {\bibinfo {title} {A silicon {B}rillouin laser},\ }\href
  {https://doi.org/10.1126/science.aar6113} {\bibfield  {journal} {\bibinfo
  {journal} {Science}\ }\textbf {\bibinfo {volume} {360}},\ \bibinfo {pages}
  {1113} (\bibinfo {year} {2018})}\BibitemShut {NoStop}%
\bibitem [{\citenamefont {Sarabalis}\ \emph {et~al.}(2016)\citenamefont
  {Sarabalis}, \citenamefont {Hill},\ and\ \citenamefont
  {Safavi-Naeini}}]{SAR16-APLP}%
  \BibitemOpen
  \bibfield  {author} {\bibinfo {author} {\bibfnamefont {C.~J.}\ \bibnamefont
  {Sarabalis}}, \bibinfo {author} {\bibfnamefont {J.~T.}\ \bibnamefont
  {Hill}},\ and\ \bibinfo {author} {\bibfnamefont {A.~H.}\ \bibnamefont
  {Safavi-Naeini}},\ }\bibfield  {title} {\bibinfo {title} {Guided acoustic and
  optical waves in silicon-on-insulator for {B}rillouin scattering and
  optomechanics},\ }\href {https://doi.org/10.1063/1.4955002} {\bibfield
  {journal} {\bibinfo  {journal} {APL Photonics}\ }\textbf {\bibinfo {volume}
  {1}},\ \bibinfo {pages} {071301} (\bibinfo {year} {2016})},\ \Eprint
  {https://arxiv.org/abs/https://doi.org/10.1063/1.4955002}
  {https://doi.org/10.1063/1.4955002} \BibitemShut {NoStop}%
\bibitem [{\citenamefont {Gundavarapu}\ \emph {et~al.}(2019)\citenamefont
  {Gundavarapu}, \citenamefont {Brodnik}, \citenamefont {Puckett},
  \citenamefont {Huffman}, \citenamefont {Bose}, \citenamefont {Behunin},
  \citenamefont {Wu}, \citenamefont {Qiu}, \citenamefont {Pinho}, \citenamefont
  {Chauhan}, \citenamefont {Nohava}, \citenamefont {Rakich}, \citenamefont
  {Nelson}, \citenamefont {Salit},\ and\ \citenamefont
  {Blumenthal}}]{GUN19-NP}%
  \BibitemOpen
  \bibfield  {author} {\bibinfo {author} {\bibfnamefont {S.}~\bibnamefont
  {Gundavarapu}}, \bibinfo {author} {\bibfnamefont {G.~M.}\ \bibnamefont
  {Brodnik}}, \bibinfo {author} {\bibfnamefont {M.}~\bibnamefont {Puckett}},
  \bibinfo {author} {\bibfnamefont {T.}~\bibnamefont {Huffman}}, \bibinfo
  {author} {\bibfnamefont {D.}~\bibnamefont {Bose}}, \bibinfo {author}
  {\bibfnamefont {R.}~\bibnamefont {Behunin}}, \bibinfo {author} {\bibfnamefont
  {J.}~\bibnamefont {Wu}}, \bibinfo {author} {\bibfnamefont {T.}~\bibnamefont
  {Qiu}}, \bibinfo {author} {\bibfnamefont {C.}~\bibnamefont {Pinho}}, \bibinfo
  {author} {\bibfnamefont {N.}~\bibnamefont {Chauhan}}, \bibinfo {author}
  {\bibfnamefont {J.}~\bibnamefont {Nohava}}, \bibinfo {author} {\bibfnamefont
  {P.~T.}\ \bibnamefont {Rakich}}, \bibinfo {author} {\bibfnamefont {K.~D.}\
  \bibnamefont {Nelson}}, \bibinfo {author} {\bibfnamefont {M.}~\bibnamefont
  {Salit}},\ and\ \bibinfo {author} {\bibfnamefont {D.~J.}\ \bibnamefont
  {Blumenthal}},\ }\bibfield  {title} {\bibinfo {title} {Sub-hertz fundamental
  linewidth photonic integrated {B}rillouin laser},\ }\href
  {https://doi.org/10.1038/s41566-018-0313-2} {\bibfield  {journal} {\bibinfo
  {journal} {Nature Photonics}\ }\textbf {\bibinfo {volume} {13}},\ \bibinfo
  {pages} {60} (\bibinfo {year} {2019})}\BibitemShut {NoStop}%
\bibitem [{\citenamefont {Gyger}\ \emph {et~al.}(2020)\citenamefont {Gyger},
  \citenamefont {Liu}, \citenamefont {Yang}, \citenamefont {He}, \citenamefont
  {Raja}, \citenamefont {Wang}, \citenamefont {Bhave}, \citenamefont
  {Kippenberg},\ and\ \citenamefont {Th\'evenaz}}]{GYG20-PRL}%
  \BibitemOpen
  \bibfield  {author} {\bibinfo {author} {\bibfnamefont {F.}~\bibnamefont
  {Gyger}}, \bibinfo {author} {\bibfnamefont {J.}~\bibnamefont {Liu}}, \bibinfo
  {author} {\bibfnamefont {F.}~\bibnamefont {Yang}}, \bibinfo {author}
  {\bibfnamefont {J.}~\bibnamefont {He}}, \bibinfo {author} {\bibfnamefont
  {A.~S.}\ \bibnamefont {Raja}}, \bibinfo {author} {\bibfnamefont {R.~N.}\
  \bibnamefont {Wang}}, \bibinfo {author} {\bibfnamefont {S.~A.}\ \bibnamefont
  {Bhave}}, \bibinfo {author} {\bibfnamefont {T.~J.}\ \bibnamefont
  {Kippenberg}},\ and\ \bibinfo {author} {\bibfnamefont {L.}~\bibnamefont
  {Th\'evenaz}},\ }\bibfield  {title} {\bibinfo {title} {Observation of
  stimulated {B}rillouin scattering in silicon nitride integrated waveguides},\
  }\href {https://doi.org/10.1103/PhysRevLett.124.013902} {\bibfield  {journal}
  {\bibinfo  {journal} {Phys. Rev. Lett.}\ }\textbf {\bibinfo {volume} {124}},\
  \bibinfo {pages} {013902} (\bibinfo {year} {2020})}\BibitemShut {NoStop}%
\bibitem [{\citenamefont {Wolff}\ \emph {et~al.}(2014)\citenamefont {Wolff},
  \citenamefont {Soref}, \citenamefont {Poulton},\ and\ \citenamefont
  {Eggleton}}]{WOL14-OE}%
  \BibitemOpen
  \bibfield  {author} {\bibinfo {author} {\bibfnamefont {C.}~\bibnamefont
  {Wolff}}, \bibinfo {author} {\bibfnamefont {R.}~\bibnamefont {Soref}},
  \bibinfo {author} {\bibfnamefont {C.}~\bibnamefont {Poulton}},\ and\ \bibinfo
  {author} {\bibfnamefont {B.}~\bibnamefont {Eggleton}},\ }\bibfield  {title}
  {\bibinfo {title} {Germanium as a material for stimulated {B}rillouin
  scattering in the mid-infrared},\ }\href
  {https://doi.org/10.1364/OE.22.030735} {\bibfield  {journal} {\bibinfo
  {journal} {Opt. Express}\ }\textbf {\bibinfo {volume} {22}},\ \bibinfo
  {pages} {30735} (\bibinfo {year} {2014})}\BibitemShut {NoStop}%
\bibitem [{\citenamefont {Liu}\ \emph {et~al.}(2017)\citenamefont {Liu},
  \citenamefont {Dostart},\ and\ \citenamefont {Popovi'c}}]{LIUARX-17}%
  \BibitemOpen
  \bibfield  {author} {\bibinfo {author} {\bibfnamefont {Y.}~\bibnamefont
  {Liu}}, \bibinfo {author} {\bibfnamefont {N.}~\bibnamefont {Dostart}},\ and\
  \bibinfo {author} {\bibfnamefont {M.~A.}\ \bibnamefont {Popovi'c}},\
  }\bibfield  {title} {\bibinfo {title} {Toward microphononic circuits on chip:
  An evaluation of components based on high-contrast evanescent confinement of
  acoustic waves},\ }\href@noop {} {\bibfield  {journal} {\bibinfo  {journal}
  {arXiv: Applied Physics}\ } (\bibinfo {year} {2017})}\BibitemShut {NoStop}%
\bibitem [{\citenamefont {Qiu}\ \emph {et~al.}(2013)\citenamefont {Qiu},
  \citenamefont {Rakich}, \citenamefont {Shin}, \citenamefont {Dong},
  \citenamefont {Solja\v{c}i\'{c}},\ and\ \citenamefont {Wang}}]{QIU13-OE}%
  \BibitemOpen
  \bibfield  {author} {\bibinfo {author} {\bibfnamefont {W.}~\bibnamefont
  {Qiu}}, \bibinfo {author} {\bibfnamefont {P.~T.}\ \bibnamefont {Rakich}},
  \bibinfo {author} {\bibfnamefont {H.}~\bibnamefont {Shin}}, \bibinfo {author}
  {\bibfnamefont {H.}~\bibnamefont {Dong}}, \bibinfo {author} {\bibfnamefont
  {M.}~\bibnamefont {Solja\v{c}i\'{c}}},\ and\ \bibinfo {author} {\bibfnamefont
  {Z.}~\bibnamefont {Wang}},\ }\bibfield  {title} {\bibinfo {title} {Stimulated
  {B}rillouin scattering in nanoscale silicon step-index waveguides: a general
  framework of selection rules and calculating sbs gain},\ }\href
  {https://doi.org/10.1364/OE.21.031402} {\bibfield  {journal} {\bibinfo
  {journal} {Opt. Express}\ }\textbf {\bibinfo {volume} {21}},\ \bibinfo
  {pages} {31402} (\bibinfo {year} {2013})}\BibitemShut {NoStop}%
\bibitem [{\citenamefont {Mirnaziry}\ \emph {et~al.}(2016)\citenamefont
  {Mirnaziry}, \citenamefont {Wolff}, \citenamefont {Steel}, \citenamefont
  {Eggleton},\ and\ \citenamefont {Poulton}}]{SAY16-OE}%
  \BibitemOpen
  \bibfield  {author} {\bibinfo {author} {\bibfnamefont {S.~R.}\ \bibnamefont
  {Mirnaziry}}, \bibinfo {author} {\bibfnamefont {C.}~\bibnamefont {Wolff}},
  \bibinfo {author} {\bibfnamefont {M.~J.}\ \bibnamefont {Steel}}, \bibinfo
  {author} {\bibfnamefont {B.~J.}\ \bibnamefont {Eggleton}},\ and\ \bibinfo
  {author} {\bibfnamefont {C.~G.}\ \bibnamefont {Poulton}},\ }\bibfield
  {title} {\bibinfo {title} {Stimulated {B}rillouin scattering in
  silicon/chalcogenide slot waveguides},\ }\href
  {https://doi.org/10.1364/OE.24.004786} {\bibfield  {journal} {\bibinfo
  {journal} {Opt. Express}\ }\textbf {\bibinfo {volume} {24}},\ \bibinfo
  {pages} {4786} (\bibinfo {year} {2016})}\BibitemShut {NoStop}%
\bibitem [{\citenamefont {Auld}(1973)}]{AUL73-BOOK}%
  \BibitemOpen
  \bibfield  {author} {\bibinfo {author} {\bibfnamefont {B.~A.}\ \bibnamefont
  {Auld}},\ }\href@noop {} {\emph {\bibinfo {title} {Acoustic fields and waves
  in solids}}}\ (\bibinfo  {publisher} {Wiley},\ \bibinfo {year}
  {1973})\BibitemShut {NoStop}%
\bibitem [{\citenamefont {Sarabalis}\ \emph {et~al.}(2017)\citenamefont
  {Sarabalis}, \citenamefont {Dahmani}, \citenamefont {Patel}, \citenamefont
  {Hill},\ and\ \citenamefont {Safavi-Naeini}}]{SAR17-OPT}%
  \BibitemOpen
  \bibfield  {author} {\bibinfo {author} {\bibfnamefont {C.~J.}\ \bibnamefont
  {Sarabalis}}, \bibinfo {author} {\bibfnamefont {Y.~D.}\ \bibnamefont
  {Dahmani}}, \bibinfo {author} {\bibfnamefont {R.~N.}\ \bibnamefont {Patel}},
  \bibinfo {author} {\bibfnamefont {J.~T.}\ \bibnamefont {Hill}},\ and\
  \bibinfo {author} {\bibfnamefont {A.~H.}\ \bibnamefont {Safavi-Naeini}},\
  }\bibfield  {title} {\bibinfo {title} {Release-free silicon-on-insulator
  cavity optomechanics},\ }\href {https://doi.org/10.1364/OPTICA.4.001147}
  {\bibfield  {journal} {\bibinfo  {journal} {Optica}\ }\textbf {\bibinfo
  {volume} {4}},\ \bibinfo {pages} {1147} (\bibinfo {year} {2017})}\BibitemShut
  {NoStop}%
\bibitem [{\citenamefont {Vogelgesang}\ \emph {et~al.}(2000)\citenamefont
  {Vogelgesang}, \citenamefont {Grimsditch},\ and\ \citenamefont
  {Wallace}}]{VOG00-APL}%
  \BibitemOpen
  \bibfield  {author} {\bibinfo {author} {\bibfnamefont {R.}~\bibnamefont
  {Vogelgesang}}, \bibinfo {author} {\bibfnamefont {M.}~\bibnamefont
  {Grimsditch}},\ and\ \bibinfo {author} {\bibfnamefont {J.~S.}\ \bibnamefont
  {Wallace}},\ }\bibfield  {title} {\bibinfo {title} {The elastic constants of
  single crystal $\beta$-si3n4},\ }\href {https://doi.org/10.1063/1.125913}
  {\bibfield  {journal} {\bibinfo  {journal} {Appl. Phys. Lett.}\ }\textbf
  {\bibinfo {volume} {76}},\ \bibinfo {pages} {982} (\bibinfo {year}
  {2000})}\BibitemShut {NoStop}%
\bibitem [{\citenamefont {Jin}\ \emph {et~al.}(2018)\citenamefont {Jin},
  \citenamefont {Polcawich}, \citenamefont {Morton},\ and\ \citenamefont
  {Bowers}}]{JIN18-OE}%
  \BibitemOpen
  \bibfield  {author} {\bibinfo {author} {\bibfnamefont {W.}~\bibnamefont
  {Jin}}, \bibinfo {author} {\bibfnamefont {R.~G.}\ \bibnamefont {Polcawich}},
  \bibinfo {author} {\bibfnamefont {P.~A.}\ \bibnamefont {Morton}},\ and\
  \bibinfo {author} {\bibfnamefont {J.~E.}\ \bibnamefont {Bowers}},\ }\bibfield
   {title} {\bibinfo {title} {Piezoelectrically tuned silicon nitride ring
  resonator},\ }\href {https://doi.org/10.1364/OE.26.003174} {\bibfield
  {journal} {\bibinfo  {journal} {Opt. Express}\ }\textbf {\bibinfo {volume}
  {26}},\ \bibinfo {pages} {3174} (\bibinfo {year} {2018})}\BibitemShut
  {NoStop}%
\bibitem [{\citenamefont {MacCabe}\ \emph {et~al.}(2019)\citenamefont
  {MacCabe}, \citenamefont {Ren}, \citenamefont {Luo}, \citenamefont {Cohen},
  \citenamefont {Zhou}, \citenamefont {Sipahigil}, \citenamefont
  {Mirhosseini},\ and\ \citenamefont {Painter}}]{MAC19-ARX}%
  \BibitemOpen
  \bibfield  {author} {\bibinfo {author} {\bibfnamefont {G.~S.}\ \bibnamefont
  {MacCabe}}, \bibinfo {author} {\bibfnamefont {H.}~\bibnamefont {Ren}},
  \bibinfo {author} {\bibfnamefont {J.}~\bibnamefont {Luo}}, \bibinfo {author}
  {\bibfnamefont {J.~D.}\ \bibnamefont {Cohen}}, \bibinfo {author}
  {\bibfnamefont {H.}~\bibnamefont {Zhou}}, \bibinfo {author} {\bibfnamefont
  {A.}~\bibnamefont {Sipahigil}}, \bibinfo {author} {\bibfnamefont
  {M.}~\bibnamefont {Mirhosseini}},\ and\ \bibinfo {author} {\bibfnamefont
  {O.}~\bibnamefont {Painter}},\ }\href@noop {} {\bibinfo {title} {Phononic
  bandgap nano-acoustic cavity with ultralong phonon lifetime}} (\bibinfo
  {year} {2019}),\ \Eprint {https://arxiv.org/abs/1901.04129} {arXiv:1901.04129
  [cond-mat.mes-hall]} \BibitemShut {NoStop}%
\bibitem [{\citenamefont {Laude}\ and\ \citenamefont
  {Beugnot}(2013)}]{LAU13-AIPA}%
  \BibitemOpen
  \bibfield  {author} {\bibinfo {author} {\bibfnamefont {V.}~\bibnamefont
  {Laude}}\ and\ \bibinfo {author} {\bibfnamefont {J.-C.}\ \bibnamefont
  {Beugnot}},\ }\bibfield  {title} {\bibinfo {title} {Generation of phonons
  from electrostriction in small-core optical waveguides},\ }\href
  {https://doi.org/10.1063/1.4801936} {\bibfield  {journal} {\bibinfo
  {journal} {AIP Advances}\ }\textbf {\bibinfo {volume} {3}},\ \bibinfo {pages}
  {042109} (\bibinfo {year} {2013})}\BibitemShut {NoStop}%
\end{thebibliography}%

\end{document}